# Model for performance prediction in multi-axis machining

*Sylvain Lavernhe (1), Christophe Tournier (1)\*, Claire Lartigue (1,2)*

*(1) Laboratoire Universitaire de Recherche en Production Automatisée*

*ENS de Cachan - Université Paris Sud 11*

*61 avenue du Président Wilson*

*94235 Cachan cedex – France*

*(2) IUT de Cachan - Université Paris Sud 11*

*9 Avenue de la division Leclerc*

*94234 Cachan Cedex – France*

*\* corresponding author :*

*tel: +33 1 47 40 29 96*

*fax: +33 1 47 40 22 20*

*Christophe.Tournier@lurpa.ens-cachan.fr*

**Abstract:**

This paper deals with a predictive model of kinematical performance in 5-axis milling within the context of High Speed Machining. Indeed, 5-axis high speed milling makes it possible to improve quality and productivity thanks to the degrees of freedom brought by the tool axis orientation. The tool axis orientation can be set efficiently in terms of productivity by considering kinematical constraints resulting from the set machine-tool/NC unit. Capacities of each axis as well as some NC unit functions can be expressed as limiting constraints. The proposed model relies on each axis displacement in the joint space of the machine-tool and predicts the most limiting axis for each trajectory segment. Thus, the calculation of the tool feedrate can be performed highlighting zones for which the programmed feedrate is not reached. This constitutes an indicator for trajectory optimization. The efficiency of the model is illustrated through examples. Finally, the model could be used for optimizing process planning.

1. INTRODUCTION

Due to a specific cutting process at high velocities, High Speed Machining (HSM) allows decreasing machining time while increasing surface quality of produced parts. The use of multi-axis machining also improves productivity. Indeed, control of the tool orientation relative to the surface reduces the number of part setups and increases the effectiveness of material removal.

The process consists of many stages, each one influencing the productivity and the final quality of the machined part (figure 1). The CAM stage calculates the tool path from the CAD model according to the interpolation format, (linear or polynomial) the driving tool direction and the CAM parameters (machining tolerance, scallop height). The choice of the parameter values directly influences the surface quality.

Following, the post-processing stage converts the calculated tool path into an adapted file for the Numerical Controller (NC), called the CNC file. The CNC file contains the set of tool positions and tool axis orientations and the corresponding feedrates. In 5-axis machining, the post-processor may also solve the Inverse Kinematical Transformation (IKT) in order to express the tool path into direct axis commands. The tool path interpolation and the trajectory follow up are thus performed by the CNC. Numerous parameters have to be managed during this step which constitutes a main difficulty [1][2]. Moreover, performance of the set "machine tool/NC unit" limits multi-axis machining benefits. For instance, during machining the actual velocity is most generally lower than the programmed one. Furthermore, velocity drops may appear [3]. As a result, machined surface quality is affected and machining time is considerably increased. As this stage, the follow-up strongly depends on the CNC parameters such as time cycles, velocity limitations and specific functions like anticipation ("look-ahead") [4][5]. During machining, performance is also altered by the axis capacities and the machine-tool architecture [6][7]. Therefore, from the CAM stage to actual machining, numerous parameters influence performance in 5-axis machining which may affect machining time as well as surface quality.

Within the framework of 5-axis machining, many methods were developed to optimize the tool axis orientation and the machining directions in order to maximize productivity while ensuring the required quality. Some methods are based on geometrical criteria to carry out this optimization such as maximization of the width of cut [8][9][10] or constant scallop height machining [11][12]. Others methods include in the optimization constraints located downstream in the machining process. For example the integration of phenomena linked to the inverse kinematical transformation during the tool path computation [13][14][15], or some constraints related to the components of the machine tool to maximize the tool feedrate [16]. In this last approach, the authors only take into account the maximum velocity of the axis motors. Acceleration, jerk and behavior of the NC unit are not considered.

From this analysis, we developed a model for performance prediction of the set "machine tool/NC unit". The objective is the prediction of trajectory portions for which slowdowns may appear in order to optimize the machining strategy. Note that our objective is not to elaborate a complete simulator of the NC unit (generally proposed by NC vendors), but to develop a model which can be easily integrated within a step of process planning optimization. Therefore, the model is simple but will efficiently replace errors based methods for finding the best machining strategy.

The model consists in three main steps (figure 2). In the first one, the calculated tool path $(X_{pr}, Y_{pr}, Z_{pr}, i, j, k)$ is transformed into a trajectory in the joint space $P^{axis1}, P^{axis2}, P^{axis3}$... by solving the IKT. Once axes are coordinated, the model enables to predict each axis' velocity profile during machining. The model integrates constraints linked to the trajectory geometry, the NC parameters (cycle times, specific functions like "look ahead") and the machine tool axis limits (maximum velocities, accelerations and jerks). The originality of this step is the use of the time inverse method (ISO 6983-1) to compare performance of translation and rotation axes. Finally, the tool feedrate is reconstructed considering the machine-tool architecture allowing the prediction of portions for which slow-downs may appear. These results may be considered to elaborate an optimal 5 axis machining strategy [16].

The paper is organized as follows. Section 2 details the parameters influencing performance of the trajectory follow-up during machining. Limits associated to these parameters are expressed as kinematical constraints. Section 3 deals with the structure of the predictive model. Kinematical constraints are expressed using the inverse time in order to generate axis velocity profiles. Section 4 is dedicated to the validation of the model through various examples. The paper is ended by some conclusions. The study relies on a five axis milling centre Mikron UCP 710 with an industrial NC Siemens 840D.

2. LIMITS OF THE FOLLOW-UP DURING MACHINING

During machining, the actual follow-up of the trajectory does not exactly match the programmed one. These differences may come from numerous sources along the process which transforms the CNC file into tool displacements: trajectory adaptation, NC performance, axis limitations, motion control regulation, deformation of the mechanical structure, tool deflection, etc...

This paper more particularly focuses on the first stage of the processing: the tool path preparation followed by the interpolation carried out by the NC unit. Due to physical and numerical limits, adaptations of the calculated tool path must be carried out during the follow-up: decrease of the relative velocity between the tool and the surface in function of each axis performance, tool path rounding according to given tolerance, etc.

Within the context of multi-axis machining, whatever the architecture of the machine-tool, there is no direct correspondence between the workpiece coordinate system and the joint space of the machine tool. Indeed, the tool path is computed in the local frame linked to the part. Then, tool positions and orientations are expressed in the joint space via the IKT in order to command axes. Hence, limits of the follow-up must be analyzed in the joint space.

### 2.1. Orders from CNC file

Information contained into the CNC file is the tool path description in the workpiece coordinate system and corresponding feedrates. Classically, the tool path is defined as a set of tool positions *(Xpr,Ypr,Zpr)* and tool axis orientations *(i,j,k)*. Corresponding axis configurations $(P^{axis1}, P^{axis2}, P^{axis3}...)$ are calculated using IKT. This calculation may lead to several solutions for axis configurations [17]. Supposing that the interpolation is linear in the joint space between two successive configurations, the trajectory followed by the tool on the surface is a curve (figure 4).

As a result, the re-sampling of the tool path in the workpiece coordinate system is necessary in order to control the deviations arising between the programmed tool path and the actual one. This step is performed in real-time by the NC unit with the constraint of respecting the Tolerance of Interpolation of the Trajectory (TIT).

In addition, axis velocities are computed from axis configurations (also called joint trajectory) taking into account the programmed feedrate. For instance, when programming the tool path in the workpiece coordinate system, the feedrate *Vf$_{prog}$* is defined as constant between two cutter location points $(Cl_1, Cl_2)$. If the length of the segment is $L_{12}$, the tool is supposed to move during $\Delta T_{12}$ from $Cl_1$ to $Cl_2$ (eq. (1)).

$$\Delta T_{12} = \frac{L_{12}}{Vf_{prog}} \quad with \quad L_{12} = \left\| \overrightarrow{Cl_1 Cl_2} \right\| \tag{1}$$

To each cutter location point and tool axis orientation programmed in the workpiece coordinate system corresponds one axis configuration $P_j$ ($P_j^{axis1}$, $P_j^{axis2}$, $P_j^{axis3}$ ...). The velocity $V_f^i$ of each axis $i$ to cover the segment $P_1^i P_2^i$ is given by eq. (2):

$$V_f^i = \frac{P_2^i - P_1^i}{\Delta T_{12}} = \frac{(P_2^i - P_1^i)}{L_{12}} \cdot Vf_{prog} \tag{2}$$

where $P_j^i$ denotes the coordinate of the joint axis $i$ for the configuration $j$. It should be noticed that if $L_{12}$ is null, equation (2) is undefined. Such a case appears when a cutter location is duplicated or when the tool axis change without any movement of the tool tip. In the first case, the duplicated point is detected and removed from the tool path. In the second case, if velocity is not specified in inverse time in the NC file, we suppose that the NC unit moves the joint axes as fast as possible while ensuring their coordination.

Finally, once the commands are expressed in the joint space, capacities of machine tool axes have to be analyzed.

### 2.2. Limits linked to machine tool axis

All trajectory long, each axis is prompted differently. Each axis behavior depends on the geometry of the joint trajectory and its discontinuities.

During linear interpolation, the follow-up of the trajectory is limited by the less powerful axis. Indeed, particularly on serial architectures, axis capacities are different. Therefore, for each elementary segment of the trajectory, the follow-up is limited by the maximum kinematical capacities (velocity, acceleration and jerk) [7] [18]. As the nature of rotation and translation movements is different, axis capacities cannot be compared directly and it becomes difficult to determine the limiting axis. To overcome this difficulty, we propose to express axis kinematical capacities using the inverse time method (see section 3).

Furthermore, discontinuities of the joint trajectory appear on block transitions. Tangency discontinuities are the most critical ones. Passing exactly through these discontinuities with a non-null feedrate would require infinite accelerations on each axis which is physically not possible. Rounding tolerances are thus introduced to improve the follow-up, while controlling the geometrical deviation to the trajectory. We develop in this section a model to adapt axis velocity to the rounding tolerance and maximum axis acceleration at tangency discontinuities. Constraints coming from this model will be included in the next section in our predictive model.

Figure 5 illustrates the kinematics of one particular axis i, in terms of position, velocity and acceleration in function of the time when a tangency discontinuity is passed with the maximal acceleration $A^i_{max}$ instead of an infinite one. The rounding error of the tool path in the workpiece coordinate system will be the result of the combination of all the machine axes.

The velocities $V^i_e$ and $V^i_b$ at the transition are reduced to satisfy at the same time the maximal deviation $e^i_{max}$ and the maximal acceleration $A^i_{max}$ allowed for each axis. Let $P^i_b$ be the point of the first linear segment defining the corner beginning, and let $P^i_e$ be the ending point of the corner belonging to the second linear segment. We make the assumption that the transition begins at $\Delta t^i/2$ before the discontinuity and ends at $\Delta t^i/2$ after. Then, the maximum variation of velocity $\Delta V^i_{max}$ between $P^i_b$ and $P^i_e$ can be defined as:

$$\Delta V^i_{max} = V^i_e - V^i_b \tag{3}$$

As the acceleration $A^i_{max}$ is assumed to be constant, the duration $\Delta t^i$ of this transition is given by:

$$\Delta t^i = \frac{\Delta V^i_{max}}{A^i_{max}} \tag{4}$$

The deviation $e^i_{max}$ between $P^i_2$ and $P^i_4$ is defined by:

$$e^i_{max} = P^i_4 - P^i_2 \tag{5}$$

On the programmed tool path, $P_2^i$ is supposed to be the end point of a linear segment traveled at a constant speed $V_b^i$. By integrating constant velocity, this yields to:

$$P_2^i = P_b^i + V_b^i \cdot \frac{\Delta t^i}{2} \tag{6}$$

$P_4^i$ is supposed to be located on the transition tool path traveled at constant acceleration $A_{max}^i$. By integrating acceleration, this yields to:

$$P_4^i = P_b^i + V_b^i \cdot \frac{\Delta t^i}{2} + \frac{1}{2} \cdot A_{max}^i \cdot \frac{\Delta t^i}{2} \tag{7}$$

Finally, we can define the deviation $e_{max}^i$ regarding to the maximum variation of velocity $\Delta V_{max}^i$ and the imposed constant acceleration $A_{max}^i$:

$$e_{max}^i = \frac{\left(V_e^i - V_b^i\right)^2}{8 \cdot A_{max}^i} = \frac{(\Delta V_{max}^i)^2}{8 \cdot A_{max}^i} \tag{8}$$

Depending on the $e_{max}^i$ value, $\Delta V_{max}^i$ will be computed in the predicive model and considered as a constraint.

### 2.3. Limits linked to the NC unit

Between two tool positions, the NC unit needs at least one interpolation cycle time to calculate axis commands [5]. Consequently, the programmed feedrate is lowered by the NC to satisfy this cycle time:

$$V_{max}^{NC} = \frac{L_{12}}{T_{cycle\,time}} \tag{9}$$

All the considerations previously exposed are supporting the predictive model of the kinematical behavior for the set "machine tool/NC unit".

3. PREDICTIVE MODEL OF KINEMATICAL BEHAVIOR

The main objective is to evaluate the actual velocity of each axis during multi-axis machining. The evolution of the velocity throughout trajectory must highlight the location of trajectory portions for which feed-rate will be strongly reduced. The analysis can be used to optimize the step of tool path computation by the CAM software. Moreover, the predictive model allows the reconstruction of the relative velocity tool-surface. It is then possible to evaluate the impact of kinematical performance on productivity as well as on geometric quality of the machined surface.

The formalism used can be considered as an extension of the programming method called inverse time (ISO 6983-1). It consists in expressing a kinematical characteristic (position, velocity, acceleration...) through its inverse time form. With such formalism it is possible to compare kinematical performance of the translational and the rotational axes. For a given trajectory, the model directly reveals which axis is the limiting one, with respect to the following characteristics: maximum velocity, maximum acceleration and maximum jerk.

### 3.1. Time inverse method

Let us consider the movement of the axis $i$ from the position $P_1^i$ to the position $P_2^i$. The axis displacement from one position to the other one is:

$$\Delta P_{12}^i = P_2^i - P_1^i \tag{10}$$

By assuming that the interpolation is linear in the joint space, the current position of the axis between the two positions is expressed as follows:

$$p_{12}^i = \Delta P_{12}^i \cdot \alpha^i \quad \alpha^i \in [0,1] \tag{11}$$

where $\alpha^i$ is the fraction of the total displacement between $P_1^i$ and $P_2^i$.

Thus, the expression of the current position of the axis in the inverse time form is:

$$\hat{p}_{12}^i = \frac{p_{12}^i}{\Delta P_{12}^i} = \alpha^i \tag{12}$$

The velocity of the axis is thus obtained by differentiation of equation (11):

$$v_{12}^i = \frac{dp_{12}^i}{dt} = \Delta P_{12}^i \frac{d\alpha^i}{dt} \qquad (13)$$

This yields to the expression of the velocity of the axis in the inverse time form:

$$\hat{v}_{12}^i = \frac{d\alpha^i}{dt} = \frac{v_{12}^i}{\Delta P_{12}^i} \qquad (14)$$

It can be noticed that $\hat{v}_{12}^i$ is equal to the inverse time which is necessary to go from configuration 1 to configuration 2.

Finally, we can express in the same manner acceleration and jerk:

$$\begin{cases} \hat{a}_{12}^i = \dfrac{d^2\alpha^i}{dt^2} = \dfrac{a_{12}^i}{\Delta P_{12}^i} \\ \hat{j}_{12}^i = \dfrac{d^3\alpha^i}{dt^3} = \dfrac{j_{12}^i}{\Delta P_{12}^i} \end{cases} \qquad (15)$$

Displacements of the axes are coordinated with respect to the joint trajectory. This coordination is implicit in inverse time since each axis displacement is reduced to a unit displacement. This involves for a joint trajectory segment:

$$\begin{cases} \hat{v}_{12}^i = \hat{v}_{12} \\ \hat{a}_{12}^i = \hat{a}_{12} \;\; \forall \; axis \; i \\ \hat{j}_{12}^i = \hat{j}_{12} \end{cases} \qquad (16)$$

For each displacement, there is a limiting axis with respect to each kinematical characteristic, velocity, acceleration and jerk. According to these limits, we determine the maximum kinematical characteristics to be respected:

$$\hat{V}_{\max 12}^{axis} = \min_i\left(\frac{V_{\max}^i}{\Delta P_{12}^i}\right) \quad ; \quad \hat{A}_{\max 12}^{axis} = \min_i\left(\frac{A_{\max}^i}{\Delta P_{12}^i}\right) \quad ; \quad \hat{J}_{\max 12}^{axis} = \min_i\left(\frac{J_{\max}^i}{\Delta P_{12}^i}\right) \qquad (17)$$

Equations (2) and (9) can be reformulated in this way considering that $\Delta P^i_{12}$ is a unit displacement in inverse time:

$$\hat{V}_{f12} = \frac{V_f^i}{\Delta P_{12}^i} = \frac{Vf_{prog}}{L_{12}} \quad ; \quad \hat{V}_{max\,12}^{NC} = \frac{V_{max}^{NC}}{L_{12}} = \frac{1}{T_{cycle\,time}} \quad (18)$$

Finally, for a segment trajectory, the velocity is limited by the minimum value of the constraints. This yields to:

$$\begin{cases} 0 \leq \hat{v}_{12} \leq \min(\hat{V}_{f12}, \hat{V}_{max\,12}^{NC}, \hat{V}_{max\,12}^{axis}) \\ -\hat{A}_{max\,12}^{axis} \leq \hat{a}_{12} \leq \hat{A}_{max\,12}^{axis} \\ -\hat{J}_{max\,12}^{axis} \leq \hat{j}_{12} \leq \hat{J}_{max\,12}^{axis} \end{cases} \quad (19)$$

### 3.2. Prediction of the velocity profiles

This step consists in determining the evolution of the position, the velocity and the acceleration of each axis in function of the time by integrating constraints previously calculated. For this purpose, the principle is the calculation of kinematical profiles in the inverse time form. These profiles are thus projected onto the trajectory.

At this stage, we must choose some parameters and functions of the NC unit. First, we choose the piloting mode of the axes by constant jerk, i.e., trapezoidal profile of acceleration. Henceforth, this piloting mode is the most popular for high-speed machines. The jerk can take the next three values:

$$\hat{j}(t) = \begin{cases} -\hat{J}_{max}^{axis} \\ 0 \\ \hat{J}_{max}^{axis} \end{cases} \quad (20)$$

After a first integration, we obtain the acceleration:

$$\hat{a}(t) = \hat{a}_0 + \hat{j}.\Delta t \quad (21)$$

with the following constraints: $-\hat{A}_{max}^{axis} \leq \hat{a}(t) \leq \hat{A}_{max}^{axis}$

A second integration gives the velocity:

$$\hat{v}(t) = \hat{v}_0 + \hat{a}.\Delta t + \frac{1}{2}\hat{j}.\Delta t^2 \quad (22)$$

with the following constraints: $0 \le \hat{v}(t) \le min(\hat{V}_f, \hat{V}_{max}^{NC}, \hat{V}_{max}^{axis})$

Then, the calculation of sampled trajectory profiles is carried out according to the frequency of the controls of the position loop (figure 6).

The last step of the velocity prediction concerns the integration of the dynamical anticipation, also called "look ahead" which allows anticipating the constraints to be respected during the trajectory follow-up [4]. This function is used to prevent overshoots. For example, if the anticipation is realised on 10 blocks; in order to generate the profile of block *N*, the geometry of the trajectory until block *N+10* is taken into account. Hence, the velocity deceleration is delayed and overshoots are avoided. The use of such a function enables to reach higher velocities (figure 7).

### 3.3. Feedrate Prediction

Actual feedrate of the tool relatively to the surface is rebuilt from each axis velocity. Cutting conditions have to be expressed on the Cutter contact point (Cc point). Equation (23) expresses the tool feedrate in the machining direction. It can be approximated by the velocity on Cl point as Cc and Cl points are close enough and rotational axis velocities are low compared to the translational ones (eq. (24) and (25)).

$$\vec{V}_{Cc, tool/surface} = \vec{V}_{Cc, tool/machine} - \vec{V}_{Cc, surf/machine} \qquad (23)$$

$$\vec{V}_{Cc, tool/machine} = \vec{V}_{Cl, tool/machine} \qquad (24)$$

$$\vec{V}_{Cc, surf/machine} = \vec{V}_{Cl, surf/machine} + \overrightarrow{CcCl} \wedge \vec{\Omega}_{surf/machine} \approx \vec{V}_{Cl, surf/machine} \qquad (25)$$

Thus, combining Cl positions, axes velocities and machining parameters, cutting conditions are evaluated along the tool path.

## 4. MODEL VALIDATION

### 4.1. Test on a single tool path

To illustrate our approach, the machining behaviour on a blending radius of 5 mm is studied (figure 8). The programmed machining strategy is parallel to plane with a toroïdal endmill (D=10mm, Rc=1mm). Chordal deviation (machining tolerance) is set to 0.01mm; tool inclination is set to 5° and feedrate to 5 m/min. 5-axis machining is carried out on a Mikron milling centre (CAXYZ structure) equipped with a Siemens Sinumerik 840D NC unit. The programming frame is oriented on the rotate table such as only the YZA axes are used.

First, the IKT is performed. Figure 9 shows calculated axis configurations corresponding to the programmed CL points. Each dot in the picture defines a position, which leads to 27 blocks.

Figure 10 compares the predicted velocities to measurements done in real time through the CNC during machining. This is an integrated functionality of the Siemens 840D.

For the first and the last blocks, where tool axis orientation does not change, predicted axis velocities are equal to the measured one. Differences appear when the rotational axis moves. The CNC treatment seems to change when commuting from two axis interpolation (YZ) to three axis interpolation (YZA). Indeed, on block 2, 3, 25 and 26 (t≈1.5 and 3 sec.), initial and final accelerations are set to zero; slowdowns appear, whereas, between blocks 4 and 24, profile is optimal. This constraint prevents axes to reach higher feedrate. Predicted axis behaviors are locally more dynamical than the actual ones. These treatment modifications by the CNC are quite difficult to predict.

Figure 11 compares the relative feedrate tool-surface calculated from measured axis velicities and predicted ones; profiles are similar. Thanks to these results evaluation of cutting conditions can be carried out; it reveals that programmed velocity is not respected on the bending radius. Indeed, as the first and last blocks are long enough, the feedrate reaches the programmed value. When the tool axis orientation varies, discontinuities created in the joint space make the feedrate falls to 0 m/min. Along the radius, the maximal velocity reached is close to 1 m/min.

To summarize, the reconstruction on the tool feedrate gives a criterion to qualify cutting conditions, and consequently the quality of the machined part

### 4.2. Test on a complex surface

The second test is performed on a complex surface. When critical portions are determined, our objective is afterwards to find the best strategy to machine the surface. This part of the problem is a not discussed in this paper [19]**.** The selected surface is a hyperbolic paraboloid (one unique Bézier patch) (figure 12). The tool path is calculated using home algorithms based on a surface representation of the trajectories.

The machining strategy used is one-way parallel planes, for which planes are oriented by 45° relatively to the surface so that trajectories correspond to the surface rules. Thus, the trajectory of a point in the workpiece coordinate system is a straight line as the tool axis is oriented with a constant angle of inclination of 5°. The programmed feedrate is 5 m/min.

The IKT is carried out in real-time by the NC unit. The authorized variations by axis are of 0.02mm for the translational axes and 0.05° for the rotational ones. The part set-up within the machine-tool workspace is such that the programming frame corresponds to the machine frame. It is important to notice that for this example, all the axes of the machine are in movement during machining.

We concentrate on a trajectory corresponding to one pass located near the centre of the surface. Simulations using the previously exposed model are illustrated in figure 13. It can be observed that, due to the velocity limits along the blocks, the programmed feedrate can never be reached. Indeed, the programmed feedrate is greater than the maximum velocity with respect to the cycle time of interpolation all pass long. Trajectory segments are thus too short to reach the programmed feedrate (equations (9) and (18)). Moreover, close to the middle of the trajectory, the maximum performances of the C axis are under other axis limits. We can conclude that C is the limiting axis.

Note that the small undulations at the beginning and at the end of the pass are due to dynamic anticipation. If one strongly increases the number of anticipated blocks, the two velocity limits can be reached.

The general shape of the predicted and the measured relative tool-surface velocities corresponds although the model overestimates the velocity on the first portion of the trajectory.

To summarize, the proposed model allows the prediction of the trajectory follow-up. In particular, it highlights the limiting axis through the evaluation of kinematical profiles (position, velocity, and acceleration). Indeed, from those profiles, trajectory portions for which the follow-up is strongly decreased. Therefore, it is possible to modify the trajectory so that the follow-up is improved. In particular, the tool axis orientation can be efficiently calculated by integrating those kinematical constraints. The trajectory optimization can be approached according the two following ways:

- A local modification of the trajectory in order to avoid slow-downs of the federate, which can be sources of marks on the part,

- A calculation of an optimal trajectory, integrating limits and constraints linked to the set machine tool/NC, in order to optimize the federate.

The final objective is to improve productivity while controlling geometrical deviations.

5. CONCLUSION

In this paper, we have presented a predictive model that evaluates axis velocities from a NC file integrated NC and axis capacities. Thanks to its specific formalism, the model can be applied whatever the machine tool architecture and the axis number. The formalism used is an extension of the inverse time method and consists in expressing each kinematical characteristic of position, velocity, and acceleration through its inverse time form. For a given trajectory such formalism allows the comparison of kinematical performances of translational and rotary axes and provides the most limiting axis with regard the trajectory follow-up. Through an example, we showed that predicted velocity profiles match the measured ones. Zones for which velocity decreases are detected by reconstruction of the relative velocity tool-surface. Nevertheless, the complexity and the specificity of industrial NC units in multi-axis machining make difficult a very sharp modeling of the kinematical behavior. However, the model is a good indicator of the actual follow-up.

These works are currently being integrated in a surface based model for the description of the tool trajectories [20]. The objective is to optimize the follow-up by a modification of the machining strategy and more particularly the tool axis orientation.

REFERENCES

[1] Y. Altintas, Manufacturing Automation. Metal Cutting Mechanics, Machine Tool Vibrations, and CNC Design, Cambridge University Press, ISBN 0-521-65973-6, 2000

[2] K. Erkorkmaz, Y. Altintas, High Speed CNC system design – Part I: Jerk limited trajectory generation and quintic spline interpolation, International Journal of Machine Tools & Manufacture, 41(9) (2001) 1323-1345

[3] A. Dugas, J-J. Lee, J-Y. Hascoët, High Speed Milling–Solid simulation and machine limits, Integrated Design and Manufacturing in Mechanical Engineering, Kluwer Academic Publishers, (2002) 287-294.

[4] Siemens, Description of functions – Sinumerik 840D/840Di/810D – Basic Machine, www.automation.siemens.com/doconweb/; 2004

[5] Siemens, Description of functions – Sinumerik 840D/840Di/810D – Special Functions, www.automation.siemens.com/doconweb/; 2004

[6] R.T. Farouki, Y-F. Tsai, C.S. Wilson, Physical constraints on feedrates and feed accelerations along curved tool paths, Computer Aided Geometric Design, 17(4) (2000) 337-359

[7] S.D. Timar, R.D. Farouki, T.S. Smith, C.L. Boyadjieff, Algorithms for time-optimal control of CNC machines along curved tool paths, Robotics and Computer-Integrated Manufacturing, 21(1) (2005) 37-53

[8] Y.-S. Lee, H. Ji, Surface interrogation and machining strip evaluation for 5-axis CNC die and mold machining, International Journal of Production Research, 35(1) (1997) 225-252

[9] C-J. Chiou, Y-S. Lee, A machining potential field approach to tool path generation for multi- axis sculptured surface machining, Computer Aided Design, 34(5) (2002) 357-371

[10] S.P. Radzevich, A closed-form solution to the problem of optimal tool-path generation for sculptured surface machining on multi-axis NC machine, Mathematical and Computer Modelling, 43(3-4) (2006) 222-243

[11] Y.-S. Lee, Non-isoparametric tool path planning by machining strip evaluation for 5-axis sculptured surface machining, Computer Aided Design, 30(7) (1998) 559-570

[12] C. Tournier, E. Duc, Iso-scallop tool path generation in 5-axis milling, The International Journal of Advanced Manufacturing Technology, 25(9-10) (2005) 867-875

[13] E.L.J. Bohez, Compensating for systematic errors in 5-axis NC machining, Computer Aided Design, 34(5) (2002), 391-403

[14] A. Affouard, E. Duc, C. Lartigue, J-M. Langeron, P. Bourdet, Avoiding 5-axis singularities using tool path deformation, International Journal of Machine Tools and Manufacture, 44(4) (2004) 415-425

[15] M. Munlin, S.S. Makhanov, E.L.J Bohez, Optimization of rotations of a five-axis milling machine near stationary points, Computer-Aided Design, 36(12) (2004) 1117-1128

[16] T. Kim, S. Sarma, Tool path generation along directions of maximum kinematic performance; a first cut at machine-optimal paths, Computer-Aided Design, 34(6) (2002) 453-468

[17] Y.H. Jung, D.W. Lee, J.S. Kim, H.S. Mok, NC post-processor for 5-axis milling machine of table-rotating/tilting type, Journal of Materials Processing Technology, 130-131 (2002) 641-646

[18] S-H. Nam, M-Y. Yang, A study on a generalized parametric interpolator with real-time jerk-limited acceleration, Computer-Aided Design, 36(1) (2004) 27-36

[19] C. Tournier, S. Lavernhe, C. Lartigue, Five-axis high speed milling optimization, 4th International Conference on Integrated Design and Production, Casablanca Morocco, 2005

[20] C. Lartigue, C. Tournier, M. Ritou, D. Dumur, High-performance NC for High-Speed Machining by means of polynomial trajectories, Annals of the CIRP, 53(1) (2004) 317-320

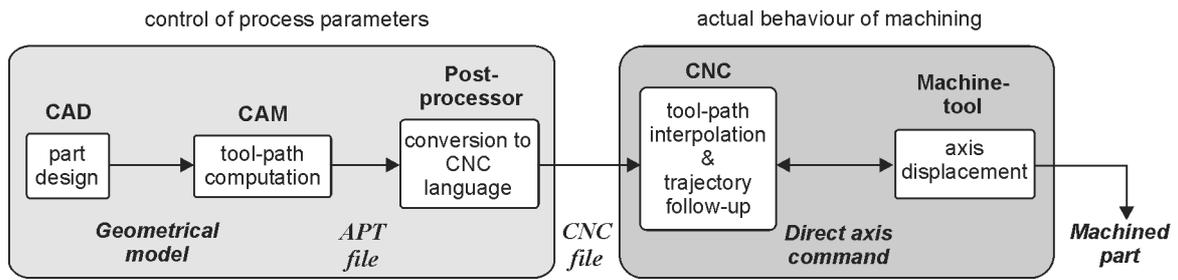

*Figure 1: Structure of the process*

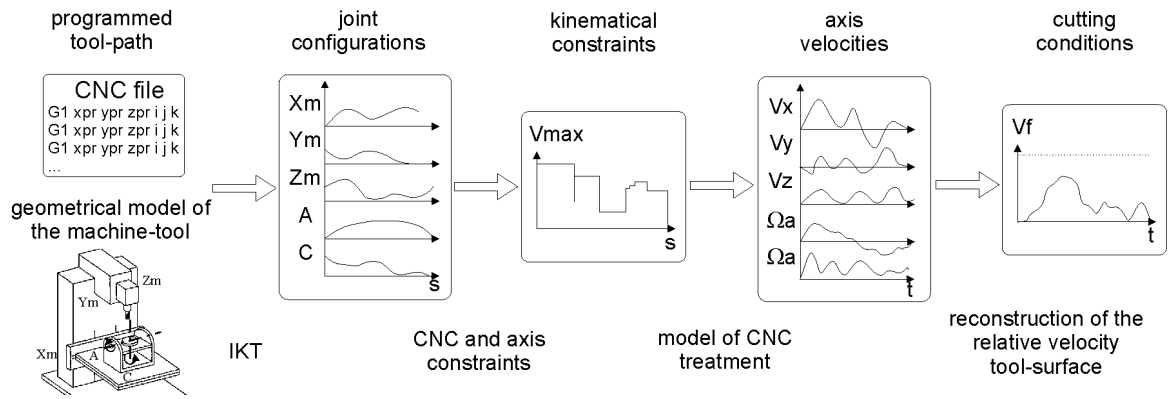

*Figure 2: Structure of the predictive model*

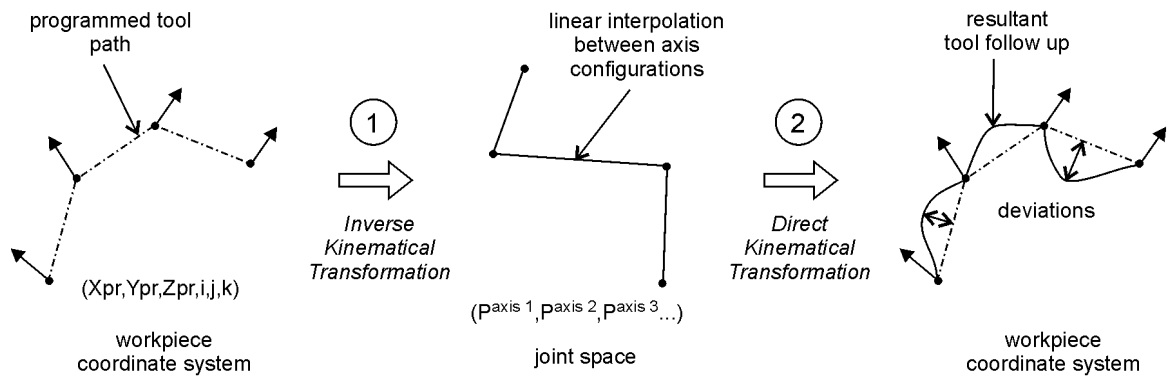

*Figure 4: Influence of the linear interpolation in the joint space on the tool path*

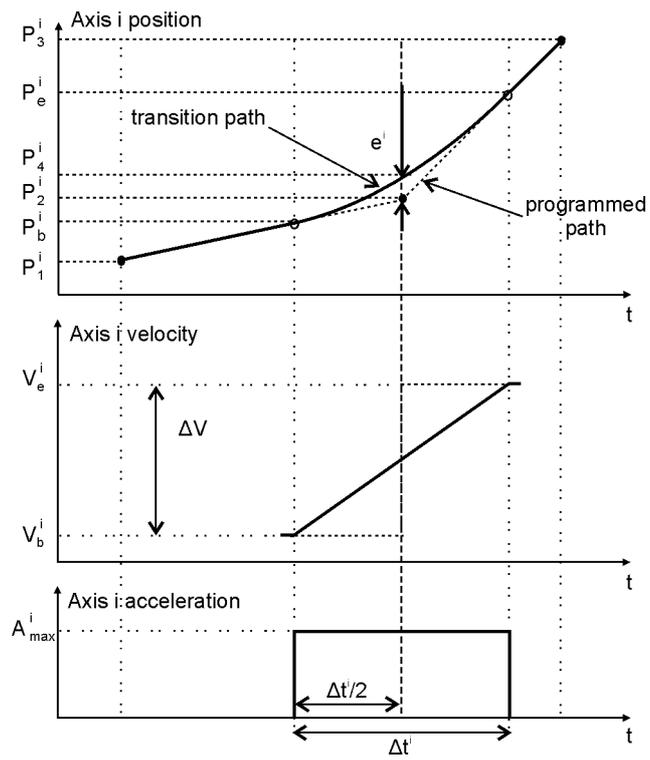

*Figure 5: Rounding tangency discontinuity*

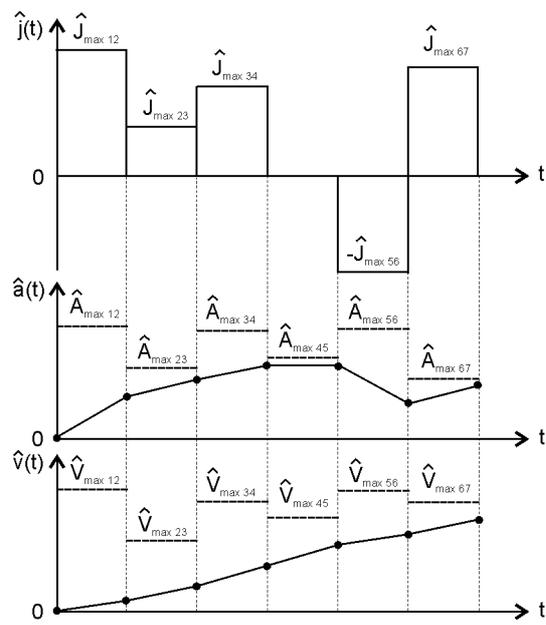

*Figure 6: Example of a jerk sequence and the resultant acceleration in inverse time*

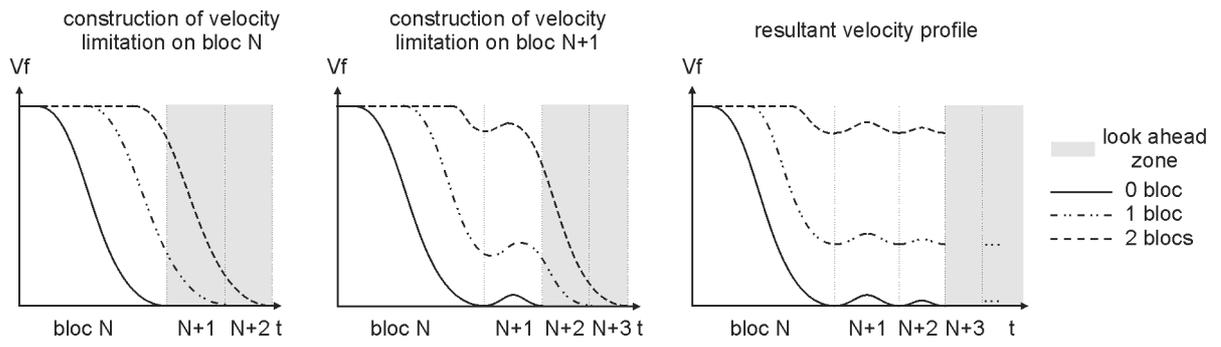

*Figure 7: Influence of look ahead on velocity profile*

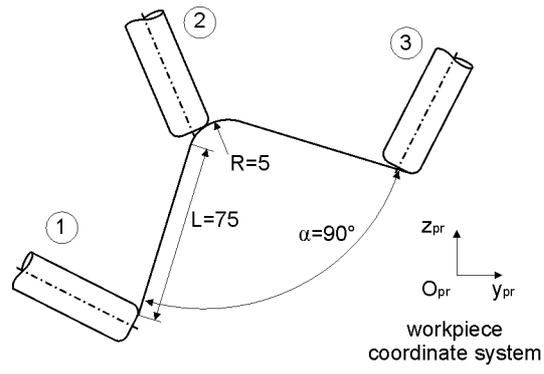

*Figure 8: Machining of the blending radius*

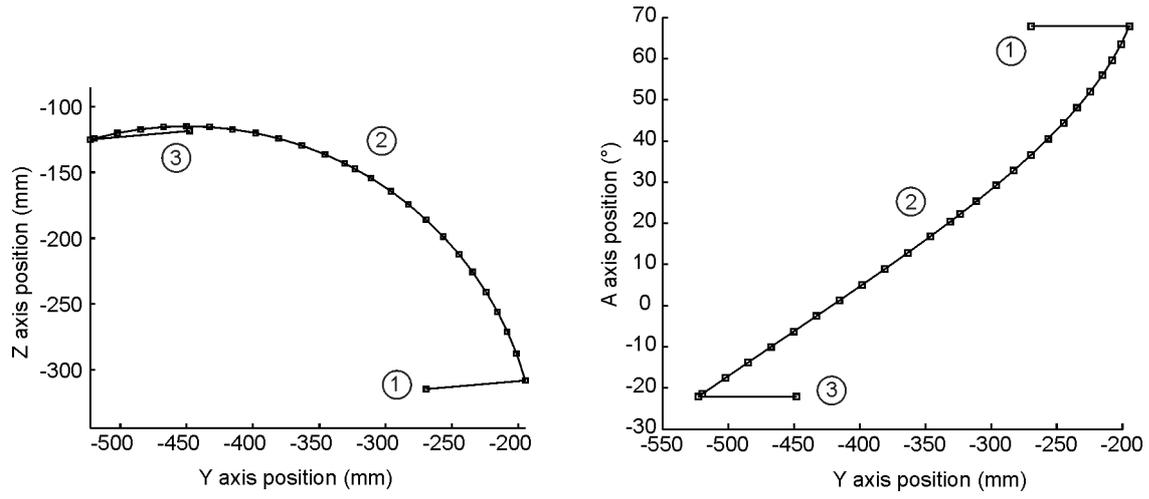

*Figure 9: Axis configurations for CL points*

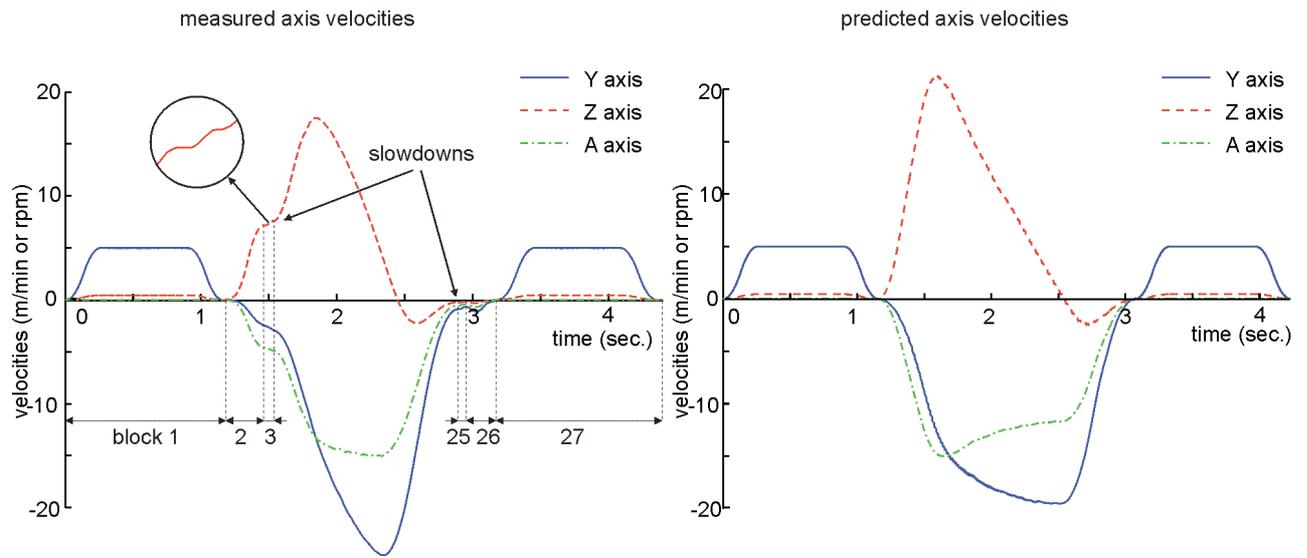

*Figure 10: Comparison between measured and predicted velocities*

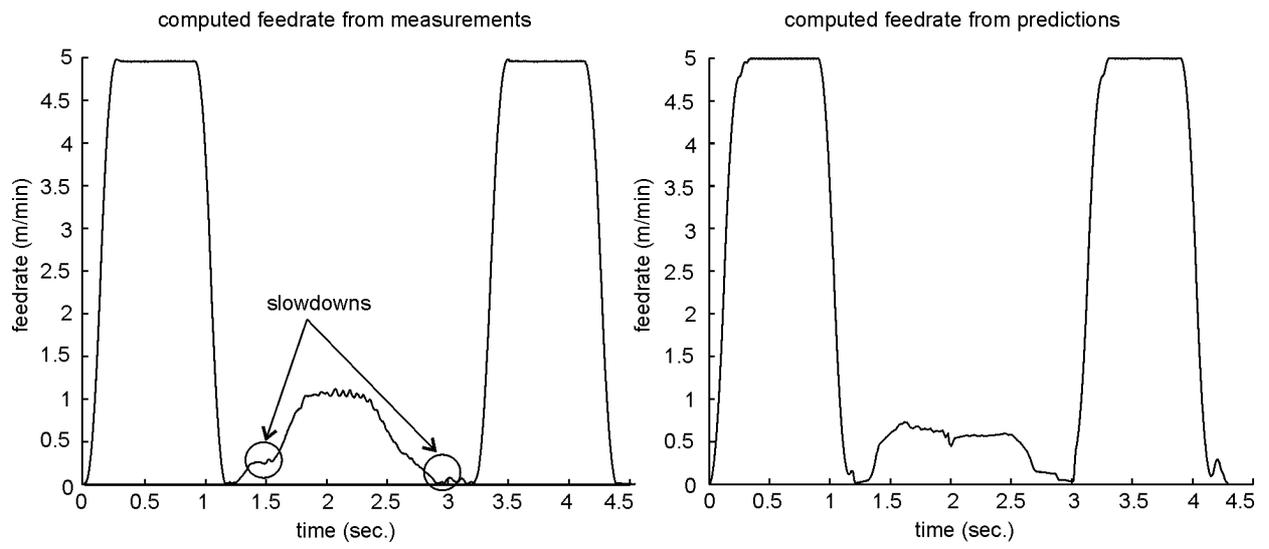

*Figure 11: Comparison between the actual feedrate and the predicted one*

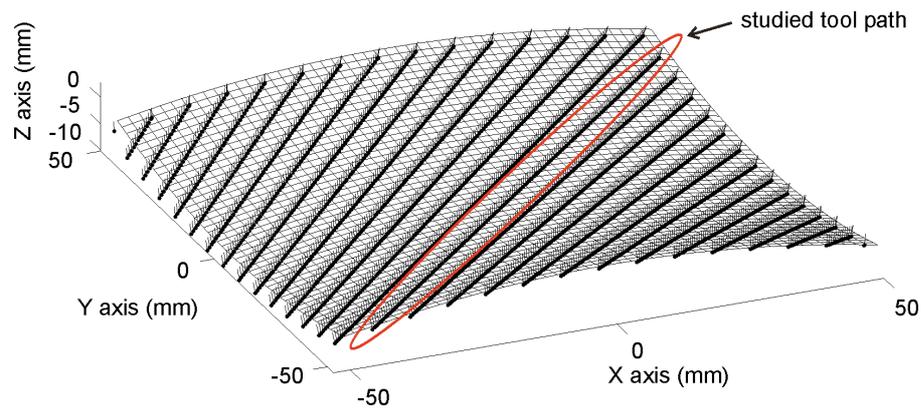

*Figure 12: Calculated tool path and part surface*

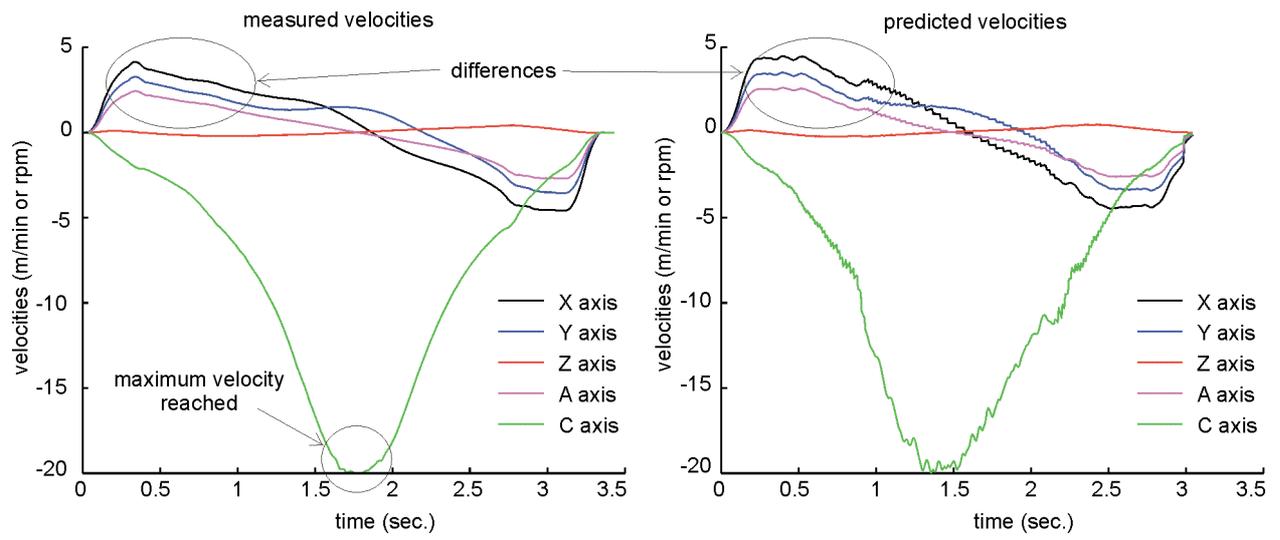

*Figure 13: Comparison between measured and predicted axis velocities*